\documentclass[prd,a4paper,showpacs,byrevtex]{revtex4}
\usepackage{graphicx}
\usepackage{dcolumn}
\usepackage{amsmath}
\usepackage{array}
\usepackage{bm}
\usepackage{amssymb}
\usepackage{amsfonts}

\begin{document}

\title[Auxiliary field method and envelope theory]
{Some equivalences between the auxiliary field method and the envelope theory}

\author{Fabien Buisseret}
\email[E-mail: ]{fabien.buisseret@umh.ac.be}
\author{Claude Semay}
\email[E-mail: ]{claude.semay@umh.ac.be}
\affiliation{Groupe de Physique Nucl\'{e}aire Th\'{e}orique, Universit\'{e}
de Mons, Acad\'{e}mie universitaire Wallonie-Bruxelles, Place du Parc 20,
B-7000 Mons, Belgium}
\author{Bernard Silvestre-Brac}
\email[E-mail: ]{silvestre@lpsc.in2p3.fr}
\affiliation{LPSC Universit\'{e} Joseph Fourier, Grenoble 1,
CNRS/IN2P3, Institut Polytechnique de Grenoble, 
Avenue des Martyrs 53, F-38026 Grenoble-Cedex, France}

\date{\today}

\begin{abstract}
The auxiliary field method has been recently proposed as an efficient technique to compute analytical approximate solutions of eigenequations in quantum mechanics. We show that the auxiliary field method is completely equivalent to the envelope theory, which is another well-known procedure to analytically solve eigenequations, although relying on different principles \textit{a priori}. This equivalence leads to a deeper understanding of both frameworks. 
\end{abstract}

\maketitle

\section{Introduction}

Finding analytical bound state solutions of the Schr\"{o}dinger equation is a problem that deserves a constant interest in mathematical physics from the beginnings of quantum mechanics. Many different techniques have been developed: perturbation theory, WKB method, variational method, etc. We refer the reader to Ref.~\cite{flu} for a complete review about analytical results in quantum mechanics. In this work, we would like to focus on two methods, both very efficient but less standard that the aforementioned techniques: the auxiliary field method (AFM) and the envelope theory (ET). 

The AFM has been first presented in Ref.~\cite{af} and extended in Ref.~\cite{af2}. It allows to find an analytical approximate solution of the eigenequation $H \left|\Psi\right\rangle=E\left|\Psi\right\rangle$, with the standard Hamiltonian
\begin{equation}\label{inco}
	H=T(\vec p^{\, 2})+V(r),
\end{equation}
provided that the eigenequation  
\begin{equation}\label{analy}
	H_A(\nu) \left|\Psi_A(\nu)\right\rangle=\left[T(\vec p^{\, 2})+\nu\, P(r)\right] \left|\Psi_A(\nu)\right\rangle= \epsilon_A(\nu)\left|\Psi_A(\nu)\right\rangle
\end{equation}
is analytically solvable, $\nu$ being a real parameter. The basic idea of the method consists in introducing an extra operator, called the auxiliary field, that transforms $V(r)$ into a function of $P(r)$ and of this auxiliary field. Analytical energy formulae can then be obtained by assuming that the auxiliary field is no longer an operator but a variational parameter. This is actually a mean field approximation with respect to the auxiliary field. The AFM has already been used to get accurate analytical energy formulae for radial power-law potentials of the form $a\, r^\lambda+b\, r^\eta$~\cite{af,af2} and for exponential potentials of the form $-\alpha\, r^\lambda\, {\rm e}^{-\beta\, r}$~\cite{af3}. Notice also that, although these last three references deal with a nonrelativistic kinetic term of the form $\vec p^{\, 2}/(2m)$, analytical formulae can also be found in the case of a relativistic kinetic term of the form $\sqrt{\vec p^{\, 2}+m^2}$~\cite{af4}. 

The ET is nearly twenty years older than the AFM and has been proposed in Ref.~\cite{env0}. The goal of this method is also to perform an analytical resolution of the eigenequation associated to the Hamiltonian~(\ref{inco}). In the ET, the potential $V(r)$ is replaced by a so-called tangential potential $V^t(r)$, depending on an extra parameter $t$ and such that $V(t+\varepsilon)-V^t(t+\varepsilon)=O(\varepsilon^2)$ for any $t$ and $\varepsilon\ll1$. The tangential potential thus provides an approximation of the exact potential at each contact point $r=t$. One can say that it generates an envelope representation of $V(r)$. Approximate analytical solutions for the eigenvalues associated with $V(r)$ are eventually known provided that $V^t(r)$ is analytically solvable. The ET has currently been applied to a wide range of problems: funnel~\cite{envpl} and Yukawa potentials~\cite{envyuk}, $N$-body systems~\cite{envnb}, relativistic Hamiltonians~\cite{env,env2}, etc.       

It has to be stressed that the AFM and the ET have been developed independently and are, \textit{a priori}, two distinct frameworks from a conceptual point of view. In order to understand the peculiarities of these two methods, we present them in Secs.~\ref{AFM} and \ref{ET}. Then, we show in Sec.~\ref{equi} that the AFM and the ET are equivalent, and we comment this equivalence.

\section{The auxiliary field method}\label{AFM}

We recall here the main points of the AFM and refer the reader to Refs.~\cite{af,af2} for more details. The problem is to find analytical approximate eigenenergies of the Hamiltonian~(\ref{inco}). The AFM suggests the following procedure. We first assume that $H_A=T(\vec p^{\, 2})+\nu\, P(r)$, where $\nu$ is a real parameter, admits bound states with an analytical spectrum. The eigenequation (\ref{analy}) is thus analytically solvable, as well as the one associated with the Hamiltonian 
\begin{eqnarray}\label{htdef}
	\tilde H(\nu)&=&T(\vec p^{\, 2})+ \tilde V(r) \nonumber \\
	&&\textrm{with}\quad \tilde V(r)=\nu\, P(r)+V\left(I(\nu)\right)-\nu\, P\left(I(\nu)\right), 
\end{eqnarray}
the function $I(x)=K^{-1}(x)$ being the inverse function of $K(x)$ defined by the relation (the prime denotes the derivative)
\begin{equation}\label{rdef}
	K(r)=\frac{V'(r)}{P'(r)}.
\end{equation}
Let us remark that $\tilde V(r)$ is of the form $C_1 P(r) + C_2$ where $C_1$ and $C_2$ are constants. These numbers must be determined in order that $\tilde V(r)$ approximates at best the potential $V(r)$.   
Using the notation defined in Eq.~(\ref{analy}), the eigenvalue $\tilde E(\nu)$ of the Hamiltonian~(\ref{htdef}) is given by
\begin{equation}\label{en1}
	\tilde E(\nu)=\epsilon_A(\nu)+V\left(I(\nu)\right)-\nu\, P\left(I(\nu)\right).
\end{equation}
The approximate eigenvalues and eigenstates of the Hamiltonian~(\ref{inco}) are eventually given by $\tilde E(\nu_0)$ and $\left|\Psi_A(\nu_0)\right\rangle$ respectively, with $\nu_0$ achieving an extremum of the total energy~(\ref{en1}), \textit{i.e.} satisfying
\begin{equation}\label{enmin}
\left.\partial_\nu \tilde E(\nu)\right|_{\nu=\nu_0}=0.
\end{equation}
The value of $\nu_0$ depends on the quantum numbers of the state considered. Once $\nu_0$ is known, the constants $C_1$ and $C_2$ can be computed.

This procedure can be justified as follows. The arbitrary potential $V(r)$ can be replaced by $\tilde V(r,\hat \nu)=\hat \nu\, P(r)+V\left( I(\hat \nu)\right)-\hat \nu\, P\left(I(\hat \nu)\right)$, a function of the so-called auxiliary field $\hat \nu$ and of an analytically solvable potential $P(r)$. The exact elimination of this auxiliary field by $\left.\delta_{\hat \nu} \tilde V(r,\hat \nu)\right|_{\hat \nu=\hat\nu_0}=0$ gives $\hat\nu_0=K(r)$ and  $\tilde V(r,\hat\nu_0)=V(r)$. Thus, the AFM would consequently lead to the exact results if the auxiliary field is seen as an operator $\hat\nu_0$. As we assume it to be a variational parameter $\nu_0$, the results are approximate but analytical. As intuitively expected, it can be checked that $\nu_0\approx \left\langle \Psi_A(\nu_0)\right|\hat\nu_0\left|\Psi_A(\nu_0)\right\rangle$, with $\hat \nu_0=K(r)$ and $\nu_0$ given by (\ref{enmin}) \cite{af}. The AFM can consequently be regarded as a ``mean field approximation" with respect to a particular auxiliary field which is introduced to simplify the calculations.

The main technical problem of the AFM is the determination of an analytical solution for the inversion of Eq.~(\ref{rdef}) and for the determination of $\nu_0$ from Eq.~(\ref{enmin}). Such a task can fortunately be achieved in many relevant cases \cite{af,af2,af3,af4}. It is of interest to mention here some general properties of the AFM~\cite{af2}:
\begin{itemize}
\item It is exact when $V(r)=P(r)$.
\item The quality of the approximation can be analytically estimated.
\item The approximate eigenvalues are ruled by the correct scaling laws.
\item The accuracy of eigenvalues can be greatly improved, being still analytical, by a comparison with exact numerical results.
\item For a potential $V(r)=\nu\, P(r)+\sigma \,v(r)$ where $\sigma$ is a small parameter and where $\sigma v(r)$ can be treated in perturbation, $\left\langle v(r)\right\rangle=v(I(\nu_0))$ at the first order in $\sigma$. 
\item Let $P_1(r)$ and $P_2(r)$ be two power-law potentials whose eigenenergies respectively read $\epsilon_1(N_1)$ and $\epsilon_2(N_2)$, where $N_1$ and $N_2$ are the principal quantum numbers. Then, the eigenenergies of $V(r)$ computed with both $P_1(r)$ and $P_2(r)$ have exactly the same functional form, but depending on $N_1$ or on $N_2$ respectively.
\end{itemize}

\section{The envelope theory}\label{ET}

As the AFM, the ET~\cite{env0} is a method aiming to get approximate 
analytical energy formulae from an arbitrary Hamiltonian~(\ref{inco}). 
We only present here its key features and refer the reader to 
Refs.~\cite{env0,env,env2,env4} for a detailed discussion about the 
basis and applications of ET.

Let us set $V(r)=v\, f(r)$ in Hamiltonian~(\ref{inco}). Then the energy 
spectrum of this Hamiltonian is formally given by $E=F(v)$, where the 
dependence on the usual quantum numbers $n$ and $l$ will be dropped for 
simplicity. The function $F(v)$ is concave but not necessarily monotone. 
This allows to define a so-called kinetic potential $k(S)$ by using the 
Legendre transformation (here 
the prime denotes the derivative with respect to $v$)
\begin{equation}\label{legendre}
    k(S)=F'(v)\quad {\rm and}\quad S=F(v)-v\, F'(v).
\end{equation}
This transformation can be understood as follows. 
$\left|\Psi\right\rangle$ being the eigenstates of 
Hamiltonian~(\ref{inco}), one can define $S=\left\langle 
\Psi\right|T(\vec p^{\, 2})\left|\Psi\right\rangle$ and rewrite formally 
the energy spectrum as $F(v)=S+v\, \left\langle 
\Psi\right|f(r)\left|\Psi\right\rangle\equiv S+v\, k(S)$. The 
transformation~(\ref{legendre}) follows from these relations. One is 
consequently led to the exact formula
\begin{equation}
    E = F(v)=\min_{S>0}\left[S+v\, k(S)\right].
\end{equation}

What can now be done to go a step further in ET is to assume that 
$V(r)=g(P(r))$, where $P(r)$ is a potential for which the solution of 
the eigenequation
\begin{equation}\label{appro}
    \left[T(\vec p^{\, 2})+v\, P(r)\right] 
\left|\Psi_A\right\rangle=\epsilon_A(v)\, \left|\Psi_A\right\rangle
\end{equation}
is analytically known. Then,  
\begin{equation}
    s=\left\langle\Psi_A\right| T(\vec p^{\, 2})\left|\Psi_A\right\rangle
\end{equation}
can be analytically computed. It can moreover be shown that the kinetic 
potential corresponding to $V(r)$, namely $K(s)$, is given approximately by
\begin{equation}
    K(s)\approx g(k_A(s)),
\end{equation}
where $k_A(s)$ is the kinetic potential associated to $P(r)$. One then 
obtains an approximate form for the eigenenergies that reads~\cite{env0}
\begin{equation}\label{ET1}
E\approx{\cal E} = \min_{s>0}\left[s+g(k_A(s))\right].    
\end{equation}
The variable $s$ actually plays the role of a variational parameter. 
But, thanks to Eq.~(\ref{appro}), the following equalities hold 
\begin{equation}
\epsilon_A(v)=s+v\, k_A(s),\quad \epsilon_A'(v)=k_A(s),
\end{equation}
and another approximate energy formula coming from the rewriting of 
Eq.~(\ref{ET1}) is
\begin{equation}\label{ET2}
    {\cal E}=\min_v 
\left[\epsilon_A(v)-v\,\epsilon_A'(v)+g(\epsilon_A'(v))\right].
\end{equation}
This last formula is called the principal envelope formula in 
Refs.~\cite{env,env2}.

It is possible to understand Eq.~(\ref{ET2}) as follows. If $V(r)=g(P(r))$, with $g(x)$ a smooth function of $x$, then we can define
the ``tangential potential" $V^t(r)$ at the point $r=t$ as 
\begin{eqnarray}\label{ET4}
	V^t(r)&=&a(t)\, P(r)+g(P(t))-a(t)\, P(t) \nonumber\\
	&& \textrm{with}\quad a(t)=\frac{V'(t)}{P'(t)}=g'(P(t)).
\end{eqnarray}
Such a particular form is obtained by demanding that $V^t(r)$ and its derivative agree with $V(r)$ and $V'(r)$ at the point of contact $r=t$. If $\varepsilon\ll 1$, one has indeed
\begin{equation}\label{vapp}
	V(t+\varepsilon)-V^t(t+\varepsilon)=\frac{ \varepsilon^2}{2}P'(t)^2\, g''(P(t))+O(\varepsilon^3).
\end{equation}
The eigenenergies of Hamiltonian $H^t=T(\vec p^{\, 2})+V^t(r)$, denoted by ${\cal E}(t)$, are given by
\begin{equation}\label{ET3}
	{\cal E}(t)=\epsilon_A(a(t))+g(P(t))-a(t)\, P(t).
\end{equation}
Let us now set 
\begin{equation}
	t=a^{-1}(v).
\end{equation}
It can be computed from Eq.~(\ref{ET4}) that $a^{-1}(v)=P^{-1}(A(v))$ with  $A(v)=g'^{-1}(v)$, and Eq.~(\ref{ET3}) becomes
\begin{equation}\label{enenv2}
	{\cal E}(v)=\epsilon_A(v)+g(A(v))-v\, A(v).
\end{equation}
The final energy spectrum has to be minimized with respect to $v$, so we have
\begin{equation}
	\left.\partial_v {\cal E}(v)\right|_{v=v_0}=0\Rightarrow A(v_0)=\epsilon'_A(v_0)
\end{equation}
and the physical energy reads 
\begin{equation}
	{\cal E}(v_0)=\epsilon_A(v_0)+g(\epsilon_A'(v_0))-v_0\, \epsilon_A'(v_0),
\end{equation}
that is nothing else than the principal envelope formula~(\ref{ET2}). 

We have just shown that ET can lead to analytical approximate energy formulae, namely Eqs.~(\ref{ET1}) and (\ref{ET2}), which are both equivalent. Moreover, Eq.~(\ref{vapp}) shows that ${\cal E}$ is a lower (upper) bound of the exact energy if the function $g$ is convex (concave), that is if $g''>0$ $(g''<0)$. The tangential potential indeed always underestimates (overestimates) the exact potential in this case. A clear advantage of ET is thus that it allows to know the variational or antivariational nature of the approximation that is performed.

\section{Equivalence between both approaches}\label{equi}

The similarity of the starting points of ET and the AFM is obvious: In both cases, a potential for which no analytical solution is known is ``approximated" by an other potential for which analytical solutions exist. It suggests that a connection between both approaches should exist; and it will indeed be established in this section. Let us apply the AFM as described in Sec.~\ref{AFM} with $V(r)=g(P(r))$. We find the following expression for the energy~(\ref{en1}) 
\begin{equation}\label{enenv}
	E(\nu)=\epsilon_A(\nu)+g\left(P(I(\nu))\right)-\nu\, P(I(\nu)),
\end{equation}
the function $I(x)=K^{-1}(x)$ being computed from the relation~(\ref{rdef}). Remarkably, this AFM formula is equal to the ET one (\ref{enenv2}) since $I(x)=P^{-1}(g'^{-1}(x))$. Consequently, the AFM and the ET lead to the same final energy formula (\ref{ET1}). The link between both approaches is given by 
\begin{equation}
	\nu=a(t).
\end{equation}
Moreover, with the point $r_0$ defined by the relation $r_0=I(\nu_0)$, the potential $\tilde V(r)$ takes the form
\begin{equation}
	\tilde V(r)=K(r_0)\, \left( P(r)-P(r_0)\right) + V(r_0). 
\end{equation}
It is then easy to see that $\tilde V(r_0)=V(r_0)$ and that $\tilde V'(r_0)=V'(r_0)$. So, the potential $\tilde V(r)$ is tangent to the potential $V(r)$. An explicit example is presented in appendix.

The function $I(x)$ can be defined if the function $K(x)$ can be inverted. In order to fulfill this condition, it is sufficient that $K(x)$ is monotonic,
that is to say that $K'(x)$ has a constant sign. But, from the definitions above, we have $K(x)=g'(P(x))$, which implies that $K'(x)=g''(P(x))\, P'(x)$. Since $K(x)$ must be monotonic, the convexity of the function $g$ is well defined if $P(x)$ is also monotonic. This is the case if $P(x)$ is a power-law potential, for instance. In these conditions, the convexity of the function $g$ can also be used to determine the variational character of the AFM. Let us note that the existence of $I(x)$ does not guarantee an analytical solution: The equation~(\ref{enmin}) must also be solved. 

Let us summarize our results. The auxiliary field $\nu$ can be introduced as an operator in the Hamiltonian~(\ref{inco}), and leads to an equivalent formulation of this Hamiltonian. If one considers it as a variational parameter rather than an operator, as in the AFM, the results are approximate but can be analytical. We have shown in this section that the auxiliary field, when seen as a variational parameter, is nothing else than the function $a(t)$ generating the tangential potential in ET. This shows that, although obtained in different ways, the AFM and the ET lead to the same results. In this way, some formulas about the power-law potentials obtained in Ref.~\cite{env3} by the ET were rediscovered with the AFM in Ref.~\cite{af}, but supplementary results are given in this last reference. 

Taking this equivalence into account, we can now better understand the meaning of the variational parameter $v$ in the ET: Its optimal value will be close to $\left\langle g'(P(r))\right\rangle$ as clearly suggested by the AFM. Moreover, the properties of the AFM that have been proven in Refs.~\cite{af,af2} also hold for ET. Finally, we can now have an \textit{a priori} knowledge of the (anti)variational nature of the AFM energy formulae provided that we express $V(r)$ as $g(P(r))$ and compute whether $g$ is convex or concave. 

\acknowledgments
C. S. and F. B. would thank the F.R.S.-FNRS for financial support. The authors thank the anonymous referee for useful suggestions.

\appendix

\section{The anharmonic oscillator}\label{example}

In order to illustrate both methods described above, they will be applied to the case of the 3D anharmonic potential $V(x)=3\, x^2+8\sqrt{\beta}\,x$ for the following dimensionless Hamiltonian
\begin{equation}
	H=\frac{\vec q^{\, 2}}{4}+3\, x^2+8\sqrt{\beta}\,x,
\end{equation}
where $\vec q$ and $\vec x$ are conjugate variables. This example is chosen because technical details can be found in Ref.~\cite{af2}.
In this paper, it is shown that, using the AFM with the potential $P(x)=x^2$, the approximate eigenenergy of $H$, with quantum numbers $n$ and $l$, is given by
\begin{equation}
\label{eq:epsapp}
\epsilon_{\textrm{app}}(\beta;n,l) = 2 \, \beta \, Y \left ( G_{-}^2(Y) + 
\frac{1}{G_{-}(Y)} \right ) \quad \textrm{with} \quad Y = \left ( \frac{N}{\beta} \right )^{2/3},
\end{equation}
where $G_{-}$ is a complicated function explicitly given in Ref.~\cite{af2} and where $N=2 n+l+3/2$. 

Following the discussion above, the ET would give exactly the same form for $\epsilon_{\textrm{app}}$. But, in addition,
with this method, it is now possible to show that the function $g(y)=3y+8\sqrt{\beta y}$ is concave ($g''(P(x))=-2 \sqrt{\beta} |x|^{-3} < 0$)
so that $\epsilon_{\textrm{app}}$ is an upper bound of the exact results. This has been checked numerically for various values of $\beta$. The envelope potential of $V(x)$ is then given by
\begin{equation}
\label{eq:venv}
\tilde V(x)=\left( 3+\frac{4\sqrt{\beta}}{x_0} \right) x^2+4\sqrt{\beta}\, x_0,
\end{equation}
where the number $x_0$ depends on the state considered. More precisely, at the minimization point, 
\begin{equation}
\label{eq:x0}
x_0=\frac{4\sqrt{\beta}}{\nu_0 -3} \quad \textrm{with} \quad \nu_0=\frac{8\, G_{-}(Y)}{Y}.
\end{equation}
It is easy to see that $\tilde V(x_0)=V(x_0)$ and that $\tilde V'(x_0)=V'(x_0)$. Moreover, the relation
\begin{equation}
\label{eq:borne}
\tilde V(x) - V(x) = \frac{4\sqrt{\beta}}{x_0} \left( x-x_0 \right)^2 \ge 0
\end{equation}
guarantees that $\epsilon_{\textrm{app}}$ is always an upper bound.

\end{document}